\documentclass[preprint]{revtex4}
\usepackage{graphicx}% Include figure files
\usepackage{dcolumn}% Align table columns on decimal point
\usepackage{bm}% bold math
\newcommand{\be}{\begin{equation}}
\newcommand{\ee}{\end{equation}}
\newcommand{\ben}{\begin{eqnarray}}
\newcommand{\een}{\end{eqnarray}}
\newcommand{\ra}{\rangle}
\newcommand{\la}{\langle}

\begin{document}
\title{Escort mean values and the characterization of power-law-decaying  probability densities}
\author{C. Tsallis$^{1,2}$\footnote{Corresponding author: tsallis@cbpf.br}, 
A.R. Plastino$^{3,4}$ and R.F. Alvarez-Estrada$^5$} 
\address{$^1$Centro Brasileiro de Persquisas Fisicas,
Xavier Sigaud 150, 22290-180 Rio de Janeiro-RJ, Brazil\\
$^2$Santa Fe Institute, 1399 Hyde Park Road, Santa Fe, NM 87501, USA \\
$^3$Physics Department, University of Pretoria, Pretoria 002,
South Africa \\  
$^4$National University La Plata,
Casilla de Correo 727, La Plata 1900, Argentina \\
$^5$Departamento de F\'{\i}sica Te\'orica I,
Facultad de Ciencias, Universidad Complutense de Madrid,
28040, Madrid, Spain
}

\date{\today}
\begin{abstract}

Escort mean values (or $q$-moments) constitute useful 
theoretical tools for describing basic features of some 
probability densities such as those which asymptotically decay like {\it power laws}. They naturally appear
in the study of many complex dynamical systems, particularly those
obeying nonextensive statistical mechanics, a current generalization 
of the Boltzmann-Gibbs theory. They recover standard mean values 
(or moments) for $q=1$. Here we discuss the characterization of a (non-negative) probability density 
by a suitable set of all its escort mean values
  together with the set of all associated  normalizing quantities, provided that all of them converge. This opens the door to a natural extension of   the well known characterization, for the $q=1$ instance, of a distribution in terms of the standard moments,  provided that {\it all} of them have {\it finite} values. 
 This question  would be  specially relevant in connection with
probability densities having {\it divergent} values for all 
nonvanishing standard moments higher than a given one (e.g., probability densities asymptotically decaying as power-laws), for which the
standard approach is not applicable. The Cauchy-Lorentz distribution, 
whose second and higher even order moments diverge, constitutes a simple 
illustration of the interest of this investigation. In this context, we also address some mathematical subtleties with the aim of clarifying some aspects of an interesting non-linear generalization of the Fourier Transform, namely, the so-called $q$-Fourier Transform.

\vskip 1cm


\end{abstract}

\maketitle

  \section{Introduction}
  
  Complex many-body systems with long-range interactions
  usually admit meta-stable states of long (but finite)
  life that eventually decay to a Boltzmann-Gibbs-like 
  state of thermodynamical equilibrium. 
  The life of
  these meta-stable states becomes longer as the size of 
  the system increases. 
  Various properties suggest that
  these meta-stable states (see \cite{PluchinoRapisardaTsallis2007,PluchinoRapisardaTsallis2008} and references therein) may be obtained from a 
  variational principle akin to the maximum entropy 
  principle associated with standard Boltzmann-Gibbs 
  thermodynamical equilibrium. 
  Along these lines, the following entropy has been introduced \cite{T88,T97,T99}
 
 \be \label{sq}
 S_q[f] \,=\, \frac{1}{q-1} \, \left(1 \,-
\, \int [f({\bf x})]^q 
  \, d\Omega \right) \,,
 \ee

 \noindent
 where $f({\bf x})$ stands for a normalized probability 
 density, ${\bf x}$ and $d\Omega$ denoting, respectively,
 a generic point and the volume element in the corresponding 
 phase space. The parameter $q$ determines the degree 
 of non-additivity exhibited by the entropic
 form (\ref{sq}).  The $q$-thermostatistical
 formalism based upon the entropic measure $S_q$ has attracted 
 considerable theoretical interest in recent years, and has led to various experimental verifications of its predictions in real physical systems: see \cite{DouglasBergaminiRenzoni2006} for cold atoms in optical lattices, \cite{LiuGoree2008} for dusty plasma, \cite{UpadhyayaRieuGlazierSawada2001} for the motion of {\it Hydra viridissima}, \cite{DanielsBeckBodenschatz2004} for defect turbulence, among others. Details can be seen in available reviews \cite{GT04,BT05,encyclopedia} and references therein. 
 Moreover,
 the $q$-thermostatistical formalism has proven
 to be a powerful theoretical tool for the treatment of
 a variegated family of problems in physics and other fields,
 ranging from the analysis of turbulence \cite{B01,B02,TB05,B07}
 and nonlinear diffusion processes \cite{PlastinoPlastino,TsallisBukman,F05,LenziMalacarneMendesPedron,NobreCuradoRowlands,SchwammleCuradoNobre,SchwammleNobreCurado} to the 
 study of economic systems \cite{Bor02}. As mentioned above, there is an increasing
 body of evidence suggesting that the probability distributions
 maximizing $S_q$ provide appropriate descriptions of meta-estable
 states in systems with long-range interactions.

  In the limit case $q\rightarrow 1$ the 
 entropic form (\ref{sq}) becomes additive and the
 standard Boltzmann-Gibbs-Shannon (BGS) entropy 

 \be \label{BGS}
 S_{BGS} \, = \, S_1 \, = - \int f({\bf x}) \ln f({\bf x}) \, d\Omega \,,
 \ee

 \noindent
 is recovered. The nonadditive character of $S_q$ is
 summarized in the relation

 \be \label{nonad}
 S_q[f^{(A+B)}] \, = \, S_q[f^{(A)}] + S_q[f^{(B)}] + 
 (1-q) S_q[f^{(A)}] S_q[f^{(B)}] \,,
 \ee

 \noindent
 where 
 $f^{(A+B)}({\bf x_A},{\bf x_B}) = f^{(A)}({\bf x_A}) f^{(B)}({\bf x_B})$ 
 is the joint probability density 
 of a composite system $A+B$ whose subsystems $A$ and $B$
 are statistically independent and described, respectively,
 by the individual probability densities $f^{(A)}$ and $f^{(B)}$.
 The third term in the right hand side of (\ref{nonad})
 corresponds to the nonadditive behaviour of $S_q$. When
 $q=1$ this term vanishes and (\ref{nonad}) reduces to the well-known
 additivity relation verified by the BGS logarithmic entropy. 

 The probability distributions obtained maximizing the measure $S_q$
 under appropriate constraints constitute the main ingredient in
 the application of the $q$-formalism to the study of specific 
 systems.  There are several theoretical reasons suggesting that
 the correct constraints to use when implementing the $S_q$
 maximum entropy principle have to be written under the form
 of escort mean values (or $q$-mean values)
 \begin{equation}
 \la A \ra_q = \frac{\int A({\bf x}) [f({\bf x})]^q \,d\Omega}{
 \int [f({\bf x})]^q \,d\Omega} \,\,. 
 \end{equation}
 In particular, 
 the quantities  $A({\bf x})$ whose mean values appear as natural 
 constraints  in many applications of the $q$-thermostatistical 
 formalism usually have divergent linear averages  $\la A \ra_1$. 
 On the contrary,  the quantities  $A({\bf x})$ provide convergent constraints if
 appropriate escort mean values are considered (more on 
 this later).  It is also worth mentioning that the entropy $S_q$, the 
 escort constraints, and the associated Lagrange multipliers 
 comply with a set of relations that have the same form as 
 the celebrated Jaynes relations \cite{J87,J05} connecting the entropy, the mean values,
 and the Lagrange multipliers appearing in the usual version
 of the maximum entropy principle \cite{TMP98}. In the particular case of 
 Gibbs' canonical distribution (and other maximum entropy distributions
 appearing in equilibrium statistical mechanics) the alluded relations reduce
 to the well known thermodynamic ones involving the system's entropy,
 the energy, temperature, and other relevant thermodynamical 
 variables \cite{TMP98}. 

 It is a well known mathematical fact that a probability 
 densitity $f(x)$ (for simplicity's sake we are going to consider only
 one dimensional situations) may be characterized by the set of 
 mean values $\la x^n \ra = \int \, x^n f(x) \, dx, \,\,\,\,\,\, 
 (n=1,2,3\ldots)$, whenever they are all finite and satisfy some restrictions \footnote{The so called {\it moment problem} in theory of probabilities is a mathematically quite complex one. The discussion is normally done separately for various classes of support of the probability distribution, namely for the $[0,1]$ support ({\it Hausdorff} moment problem), the $[0,\infty)$ support ({\it Stieltjes} moment problem), and the $(-\infty,\infty)$ support ({\it Hamburger} moment problem). General necessary and sufficient conditions are still ellusive \cite{math1,math2,math3,math4}.}. A usual way to see this is
 by recourse to the Fourier transform of $f(x)$: the 
 moment $\la x^n \ra$ is given by the $n$-th derivative 
 of the Fourier transform $F(\xi)$ of $f(x)$ (evaluated
 at $\xi =0$). Due to the important role played by escort mean 
 values in the $q$-statistical formalism and in many of its 
 applications, it is of considerable interest to explore
 possible extensions of the above characterization of 
 probability densities to scenarios involving densities which asymptotically decay as {\it power laws}. The aim of the present note is to address this 
 problem. We shall use the $q$-generalization of the Fourier transform of $f(x)$, and discuss the uniqueness of its inverse, and the intimately connected problem of  whether a probability density
 could  in general be completely determined by an appropriate set 
 of escort mean values, whenever these are {\it all} finite. The latter  condition is considerably less restrictive than demanding that {\it all} the standard mean values be finite. It is thus at this point that we open the door in the sense of generalizing the usual theorems (recovered as the $q=1$ particular case of the present study) related to the moment problem.

\section{Escort $q$-Averages and the Characterization of
probability Densities}

Let $f(x)$ be a properly normalized probability
density defined on the (one dimensional) variable $x$,

\be  \label{norm}
\int^{+\infty}_{-\infty} \, f(x) \, dx \, = \, 1 \,.
\ee

\noindent
The {\it unnormalized} $q$-moments of $f(x)$ are defined as

\be   \label{unnormalized}
\int^{+\infty}_{-\infty} x^n \, [f(x)]^q \, dx \,.
\ee

\noindent
On the other hand, the {\it normalized}
$q$-averages (also known as escort mean values)
of a given quantity $A(x)$ are

\be
\langle 
A(x)
\rangle_q \, = \,  
\int^{+\infty}_{-\infty}\, A(x) \, f_q(x) \, dx \,,
\ee

\noindent
where $f_q(x)$ stands for the {\it escort probability
density} \cite{BS93,Abe2003}, defined as,

\be \label{escort1}
f_q(x) \, = \, \frac{[f(x)]^q}{\nu_q[f]} \,,
\ee

\noindent
where

\be \label{escort2}
\nu_q [f] \, = \, 
\int^{+\infty}_{-\infty} \, [f(x)]^q \, dx \,. 
\ee

Our main instrument in order to elucidate if (and how)
a probability density can be fully determined by a 
set of escort mean values is the $q$-Fourier transform. 
The $q$-Fourier transform of a normalized (non-negative ) probability
density $f(x)$ is defined as \cite{UTS}

\be
F_q[f](\xi) \, = \, 
\int_{-\infty}^{+\infty} \, 
dx \, e_q\left( i\xi x[f(x)]^{q-1} \right)
\, f(x)\,, \,\,\,\,\, (q \ge 1) \,,
\ee
 We recall that for real x: 
\be
e_q(x) \equiv e_q^x \equiv [1+(1-q)\, x]_+^{\frac{1}{1-q}} \;\;\; (e_1^x=e^x) \,,
\ee 
where $[z]_+=z$ if $z \ge 0$, and vanishes if $z<0$. Noticing  that   an  imaginary argument is needed in  the $q$-Fourier transform, we    write the latter as  
\be\label{FqfUTS0}
F_q[f](\xi) \, = \, 
\int_{-\infty}^{+\infty} \, 
dx \,[1-(q-1)i\xi x[f(x)]^{q-1}]^{\frac{1}{1-q}} 
\, f(x)\,, \,\,\,\,\, (q \ge 1) \,.
\ee

By taking the principal value of $[1-(q-1)i\xi x[f(x)]^{q-1}]^{\frac{1}{1-q}}$,  Eq.  (\ref{FqfUTS0}) can also  be recast as \cite{UTS}: 
\be\label{FqfUTS}
F_q[f](\xi) \, = \, 
\int_{-\infty}^{+\infty} \, 
dx \, \left( 1+(q-1)^2\rho^{2}\right)^{\frac{1}{2(1-q)}}\times\exp\left(\frac{i\arctan[(q-1)\rho]}{q-1}\right)
\, f(x), \,\,\,\,\, (q \ge 1) \,,
\ee
with  $\rho\equiv\xi x[f(x)]^{q-1}$.
\noindent
\noindent
It can be verified that the derivatives of 
the $q$-Fourier transform $F_q[f](\xi)$
are closely related to an appropriate set of
unnormalized $q$-moments of the original 
probability density.
Indeed, the first few low-order derivatives
(including the zeroth order) are given by

\be
F_q[f](\xi=0) \, = \, 1 \,, 
\ee

\be
\left[ 
\frac{dF_q[f](\xi)}{d\xi} 
\right]_{\xi=0} \, = \,
i \int_{-\infty}^{+\infty} \, 
dx \, x \, [f(x)]^q \,, 
\ee

\be
\left[ 
\frac{d^2F_q[f](\xi)}{d\xi^2} 
\right]_{\xi=0} \, = \,
-q \int_{-\infty}^{+\infty} \, 
dx \, x^2 \, [f(x)]^{2q-1} \,, 
\ee

\noindent and

\be
\left[ 
\frac{d^3F_q[f](\xi)}{d\xi^3 }
\right]_{\xi=0} \, = \,
-i q(2q-1)\int_{-\infty}^{+\infty} \, 
dx \, x^3 \, [f(x)]^{3q-2} \,. 
\ee

\noindent
The general $n$-derivative is

\be \label{teoremita1}
\left[ 
\frac{d^{(n)}F_q[f](\xi)}{d\xi^n} 
\right]_{\xi=0} \, = \,
i^n 
\left\{
\prod_{m=0}^{n-1}[1+m(q-1)]
\right\}
\int_{-\infty}^{+\infty} \, 
dx \, x^n \, [f(x)]^{1+n(q-1)}, \,\,\,\,(n=1,2,3,\ldots) \,.  
\ee

\noindent
Recalling (\ref{unnormalized}), this last relation can be re-cast in terms of 
normalized $q$-mean moments $\langle x^n\rangle_q$,

\be \label{teoremita2}
\frac{1}{\nu_{q_n}}
\left[ 
\frac{d^{(n)}F_q[f](\xi)}{d\xi^n} 
\right]_{\xi=0} \, = \,
i^n 
\left\{
\prod_{m=0}^{n-1}[1+m(q-1)]
\right\}
\langle x^n \rangle_{q_n}, \,\,\,\,(n=1,2,3,\ldots) \,,  
\ee

\noindent
with

\be \label{quene}
q_n \, = \, 1+n(q-1) \,.
\ee

\noindent
Now, the derivatives (\ref{teoremita1}) determine the form
of the $q$-Fourier transform $F_q[f](\xi)$ through its Taylor 
expansion around $\xi=0$, i.e.,
\be  \label{qFTsex}
F_q[f](\xi)=1+ \left[ 
\frac{dF_q[f](\xi)}{d\xi} 
\right]_{\xi=0} \xi + \frac{1}{2}\left[ 
\frac{d^2F_q[f](\xi)}{d\xi^2} 
\right]_{\xi=0} \xi^2 + \frac{1}{3!}\left[ 
\frac{d^3F_q[f](\xi)}{d\xi^3 }
\right]_{\xi=0} \xi^3+ ...
\ee

 We shall address two related questions, namely, whether  the inverse $q$-Fourier 
transform of $F_q[f](\xi)$ (that is, the probability density $f(x)$)
is uniquely and completely determined \cite{UTS} by $F_q[f](\xi)$ (see also \cite{UmarovTsallis2007}), and whether  {\it the set of quantities $\nu_{q_n}$ and $\la x^n \ra_{q_n}$
do characterize completely the probability density $f(x)$}. Appendixes A and B will be devoted to these problems.

Naturally, Eq. (\ref{quene}) immediately leads to the following generalized escort distributions

\be \label{escort3}
f_{q_n}(x) \, = \, \frac{[f(x)]^{1+n(q-1)}}{\nu_{q_n}[f]}  \;\;\;\;(n=0,1,2,...) \,,
\ee

\noindent
where

\be \label{escort4}
\nu_{q_n} [f] \, = \, 
\int^{+\infty}_{-\infty} \, [f(x)]^{1+n(q-1)} \, dx \,,
\ee
of which the escort distribution (\ref{escort1}-\ref{escort2}) is but the $n=1$ member.

Notice a strong property, namely that $\la x^n \ra_{q_n}$ ($n=0,1,2,...$) are simultaneouly {\it all  finite} for $q<2$, and {\it all divergent} for $q\ge 2$, if $f(x)$ decays like $x^{1/(q-1)}$ (which, remarkably enough, is {\it precisely} what occurs in $q$-statistics, where $f(x) \propto e_q^{-\beta x}$). Notice also that, from Eq. (\ref{quene}), (i) $q=1$ implies $q_n=1, \; \forall n$, thus recovering as a particular case the standard theorem about characterization of a probability density through its infinite moments; (ii) $q_1=q, \; \forall q\ge 1$, thus recovering, as another interesting particular case, the form of constraints currently used in nonextensive statistical mechanics \cite{TMP98}.

We now consider the typical situation arising to complex systems such as many-body 
problems with long-range interactions and/or quantum entanglement, edge of chaos, free-scale networks, and others (all of them being, in fact, systems typically addressed through $q$-thermostatistics). Usually one has probability
densities behaving asymptotically as power laws,

\be \label{powerlaw}
f(x)  \sim |x|^{-\gamma} \,\,\,\,\, (|x| \to\infty; \,\gamma >0) \,.
\ee

\noindent
It is easy to realize that (if $f(x)$ is defined on an 
unbounded $x$-interval) the standard linear moments
$x^n$ will not be convergent for arbitrary  values
of $n$. Consequently, the standard way of characterizing
the probability density via its linear moments is not
feasible. On the other hand, let us see what happens with
the set of escort mean values appearing in equations
(\ref{teoremita1}-\ref{teoremita2}). The normalizability
of $f(x)$, and the convergence of the integrals 
defining the quantities $\nu_{q_n}$ and the unnormalized 
$q_n$-moments require, respectively, that the following relations hold,

\ben \label{normalizables}
1 &-& \gamma \, < \, 0 \,, \cr
1 &-& \gamma q_n = 1 - \gamma -n\gamma (q-1) \, <  \, 0 \,, \cr 
1 &+& n - \gamma q_n = 1 - \gamma +n[1-\gamma (q-1)] \, < \, 0 \,.
\een

\noindent
The above relations are verified provided that $\lambda $
and $q$ comply with

\be \label{normaf}
\gamma >1 \,,
\ee

\noindent
and

\be \label{normoment}
q \, \ge  \, 1 + \frac{1}{\gamma} \,\,.
\ee

\noindent
Equation (\ref{normaf}) can be assumed to hold, because it
is just the condition required for the power-like density $f(x)$ 
to be normalizable. A physically interesting class of normalizable, power-like probability densities
$f(x) \sim |x|^{-\gamma}$ (like the $q$-Gaussians \cite{UmarovTsallis2007}) can be, if some suitable conditions are satisfied, characterized by an appropriate set of
convergent escort mean values $\la x^n \ra_{q_n}$, as prescribed 
by equations (\ref{teoremita1}-\ref{quene}), provided that $q$
verifies the inequality (\ref{normoment}). We shall from now on use the most stringent value of $q$, namely
\be \label{connection}
q \, =  \, 1 + \frac{1}{\gamma}\,,
\ee
which, as already mentioned, is consistent with $q$-statistics.

 The above considerations can be nicely illustrated in the case of 
 an important family of probability distributions appearing
 in many applications of the $q$-thermostatistical theory 
 (see, for instance, \cite{BT05,PlastinoPlastino,TsallisBukman,F05} and references therein), namely 
 the $Q$-Gaussians (to avoid confusion, we adopt here the notation $Q$-Gaussians, instead of $q$-Gaussians as usually done in the literature)

 \be \label{qgau}
 G_Q(\beta,x) \, = \, \frac{\sqrt{\beta}}{C_Q} \, e_Q^{-\beta  \, x^2} \,,
 \ee

 \noindent
 which are defined in terms of the $Q$-exponential function, which, as indicated previously, satisfies
 $e_Q^{\, x} \, \equiv \, [1+(1-Q)\,x]^{\frac{1}{1-Q}}_+$.
 
 In Eq. (\ref{qgau}), $\beta $ is a positive parameter whose inverse ($1/\beta$)
 characterizes the ``width" of the $Q$-Gaussian, and $C_Q$ is 
 an appropriate normalization constant. The $Q$-Gaussians 
 constitute simple but important examples of maximum
 $q$-entropy ($q$-maxent, for short) distributions.
 The probability density $G_Q(\beta, x)$ maximizes the 
 entropy $S_Q$ under the constraints imposed by 
 normalization and the escort mean value $\la x^2 \ra_Q$.
 The parameter $\beta$ is related to the Lagrange 
 multiplier associated with the $\la x^2 \ra_Q$ constraint.
 The $Q$-Gaussian may be regarded as a paradigmatic 
 example of a $q$-maxent probability distribution.
 The probability density $G_Q(\beta, x)$ reduces, 
 of course, to a standard Gaussian distribution in 
 the limit $Q\rightarrow 1$, and recovers the Cauchy-Lorentz distribution $G_2(\beta, x) \propto 1/(1+ \beta x^2)$ for $Q=2$. The distributions (\ref{qgau}) are normalizable for $Q<3$ (the support is bounded for $Q<1$ and infinite for $1 \le Q <3$). Their second moment is {\it finite} for $Q<5/3$, and {\it diverges} for $5/3 \le Q<3$. But, their second $Q$-moment is {\it finite} for $Q<3$, hence {\it both the norm and the second $Q$-moment are mathematically well defined up to the same value of $Q$}.
 
 Now, for $Q$-Gaussians we have, using Eq. (\ref{qgau}), $G_Q(\beta,x) \propto 1/|x|^{2/(Q-1)}$ ($|x| \to\infty$), hence $\gamma = 2/(Q-1)$, with $Q>1$. 
 Consequently, for normalizable $Q$-Gaussians (i.e., $Q<3$) the 
 representation (\ref{teoremita1}-\ref{teoremita2}) can always 
 be implemented. Since, using Eq. (\ref{connection}), $\gamma=1/(q-1)$, we have 
 \be \label{connection2}
 q-1=\frac{Q-1}{2} \,,
 \ee
 hence, using Eq. (\ref{quene}), $q_n=1+n(Q-1)/2$, and therefore $q_2=Q$. This outcome precisely coincides with the well known recipe for $Q$-Gaussians whenever obtained from the optimization of $S_Q$ with fixed and finite $\la x^2 \ra_Q$ ! Consistently, we verify from Eq. (\ref{connection2}), that the well known upper bound $Q=3$ for $Q$-Gaussians, coincides with the upper bound $q=2$ for the present theory (and $q$-statistics).

\section{Conclusions}

We have argued that the  $q$-Fourier transform, which is a crucial tool for these studies,  has not a unique inverse, in general. Intimately connected to that,  we have argued also that a (non-negative ) probability density $f(x)$ cannot  be in general 
fully characterized by the set of all escort mean values 
$\la x^n \ra_{q_n}$ together with the  set of all associated quantities 
$\nu_{q_n}$, which are the integrals of the
$q_n$-powers of the density $f(x)$. However, for specific classes of inverses, depending typically on a set of generic coefficients, the use of the set  $\{\nu_{q_n}\}$, together with all the $q$-moments, is expected to be sufficient for uniquely determining the physically appropriate inverse.
It is of course required that all those escort mean values and all $\nu_{q_n}$'s converge.  Appendixes A and B deal with these mathematical subtleties.

 The exponents 
$q_n$ are given by $q_n= 1+n(q-1)$. For the important
case of power-like probability densities (i.e., $f(x)$ decaying like $1/|x|^\gamma$ for $|x| \to\infty$, with $\gamma>1$) we have 
determined the range of $q$-values (inequality 
(\ref{normoment})) for which all the alluded quantities 
are finite. The particular case $q=1$ recovers the usual connection (applicable only to distributions such that {\it all} the standard moments are {\it finite}). Making the choice $\gamma=1/(q-1)$, the whole construction is mathematically admissible for $q<2$, in full consistency with the $q$-exponential distribution proportional to $e_q^{- \beta x}$, naturally emerging within nonextensive statistical mechanics. In other words, 
although this connection implies  the use of auxiliary conditions and is subject to mathematical subtleties, it is completely independent from nonextensive statistics. In fact, it enriches the current use \cite{TMP98,Abe2003} of escort distributions in the definition of the constraints under which the entropy $S_q$ is to be optimized, to obtain the stationary-state distribution.   

In the present contribution we have considered 
representations of one dimensional probability 
densities in terms of escort mean values of
powers of the state variable $x$. It would be interesting to extend this approach to higher dimensional
situations, and to consider escort mean values 
associated with other functions of the state variables,
such as polynomials. These extensions may be useful
for the study of time dependent processes in complex systems
(e.g., systems with long-range interactions) 
using hierarchies of evolution equations derived from 
the corresponding Liouville equation (see, 
for instance, \cite{AE02,AE07}). These lines of inquiry 
will be addressed in a future publication.

Let us finally point out that  further mathematical investigations would be interesting regarding  (i)  the precise radius of convergence of the  expansion (\ref{qFTsex}) (it is nevertheless already clear that this radius is not zero, since it contains the $q$-Gaussian distributions \cite{UmarovTsallis2007,TsaQue}); and (ii) a more  extended analysis about  the precise classes of functions for which the $q$-Fourier Transform   is either invertible or  non-invertible, and about  the precise class of  probability densities $f(x)$ which are uniquely determined by the set of all    escort mean values 
$\la x^n \ra_{q_n}$'s  together with the set of all 
$\nu_{q_n}$'s.

\acknowledgements
One of us (C. T.) has benefited from interesting related conversations with H.J. Hilhorst, S. Umarov and E.M.F. Curado, and acknowledges partial financial support by CNPq and Faperj (Brazilian agencies). Another one (R.F. A.-E.) acknowledges the financial support of Ministerio de Educacion y  Ciencia (Project FPA2004-02602), Spain.

\section*{Appendix A: Non-Uniqueness of the  Inverse to the  $q$-Fourier Transform}

 Is the inverse of   $F_q[f](\xi)$, that is, the probability density $f(x)$,
 uniquely and completely determined?  \cite{UTS,UmarovTsallis2007}.
 We  shall treat in this Appendix  the issue of the uniqueness versus non-uniqueness of the  $q$-Fourier Transform.  A first argument in 1), by regarding   $(q-1)$ sufficiently small and linearity in $f$, would seem to indicate uniqueness. However, we 
 present in 2) and 3)  a  mathematical framework for non-small $(q-1)$ and including the nonlinearity in $f$, based upon analytic functions,  which allow for  local  non-uniqueness. 

1) We shall work here with  (\ref{FqfUTS}).  $ f(x)$ is supposed not to be wildly divergent at any finite $x$ and to have some power-law decay for large $\mid x \mid$. All that is required if $ f(x)$ is in the $L_{1}(R)$ class  \cite{UTS}.
We shall consider $q-1$ adequately small, expand  (\ref{FqfUTS}) into powers of $(q-1)$ and keep only orders zeroth and first, neglecting orders $(q-1)^{n}$, $n\geq 2$: 
\be\label{FqfUTS1}
F_q[f](\xi) \, \simeq  \, 
\int_{-\infty}^{+\infty} \, 
dx \, \left(1+\frac{(1-q)x^{2}\xi^{2}}{2}\right)\exp\left(ix\xi\right)
\, f(x) \,\,\,,
\ee
Then, $F_q[f](\xi)$ in (\ref{FqfUTS1}) becomes linear in $f$, but it is   more complicated than just the standard Fourier transform: in other words, (\ref{FqfUTS1}) stands midway between  the $q$-Fourier transform and the standard one. Such a linearity enables to recast  the uniqueness problem of the $q$-FT as follows. Suppose that, assuming the approximation(\ref{FqfUTS1}),  two functions  $f_{1}(x)$ and $f_{2}(x)$ have the same $q$-FT. If  $f_{2}(x)-f_{1}(x)=\epsilon (x)$, (\ref{FqfUTS1}) yields: 
\be
0 \, \simeq  \, 
\int_{-\infty}^{+\infty} \, 
dx \, \left(1+\frac{(1-q)x^{2}\xi^{2}}{2}\right)\exp\left(ix\xi\right)
\, \epsilon(x) \,\,\,,
\ee
Then, $\tilde{\epsilon}=\tilde{\epsilon}(\xi)$, the ordinary Fourier transform   of $\epsilon (x)$, fulfills:
\be\label{FqfUTS2}
0 \, \simeq  \,  \left(1+\frac{(q-1)\xi^{2}}{2}\frac{d^{2}}{d\xi^{2}}\right)
 \,\tilde{\epsilon} 
 \,\,\,,
\ee 

The expected properties of $f(x)$, stated above, suggest the following. $\tilde{\epsilon}(\xi) $ is supposed not to be wildly divergent at any finite $\xi$ and to have some power-law decay for large $\mid \xi \mid$. 

We look for  exact solutions of the   ordinary linear   second order differential equation (\ref{FqfUTS2}) bearing ing the structure $\tilde{\epsilon}(\xi)= \xi^{-\alpha}$. We readily get: $(q-1)\alpha(\alpha+1)=-2$. Then, the two linearly independent exact solutions of  (\ref{FqfUTS2}) are: 
\be
\tilde{\epsilon}(\xi)_{+} \, =  \, \xi^{\frac{1-i(-1+8/(q-1))^{1/2}}{2}}\, 
 \,\,\,\,,
\ee
\be
\tilde{\epsilon}(\xi)_{-} \, =  \, \xi^{\frac{1+i(-1+8/(q-1))^{1/2}}{2}}\, 
 \,\,\,,
\ee
Neither $\tilde{\epsilon}(\xi)_{+} $ nor  $\tilde{\epsilon}(\xi)_{+} $ have any sort of  power-law decay for large $\mid \xi \mid$. It seems reasonable to reject them. Since they are the only solutions that we have found  to  orders zeroth and first in $(q-1)$, it seems not unreasonable to infer that there are no acceptable functions $\tilde{\epsilon}(\xi)$.

We shall now outline an essentially equivalent argument. We apply $\int_{-\infty}^{+\infty}d\xi\exp\left(iy\xi\right)$ to (\ref{FqfUTS1}): 
\be\label{FqfUTS3}
0 \, \simeq  \, 
\int_{-\infty}^{+\infty}d\xi\exp\left(iy\xi\right)\int_{-\infty}^{+\infty} \, 
dx \, \left(1+\frac{(1-q)x^{2}\xi^{2}}{2}\right)\exp\left(ix\xi\right)
\, \epsilon(x) \,\,.
\ee
Eq. (\ref{FqfUTS3}) implies that  $\epsilon(y)$ fulfills:
\be\label{FqfUTS4}
0 \, \simeq  \,  \left(1+\frac{(q-1)}{2}\frac{d^{2}}{dy^{2}}\right)
 \,[y^{2}\epsilon(y)] 
 \,\,\,.
\ee 
 The exact solutions of (\ref{FqfUTS4}) have the structure $\epsilon(y)= y^{\beta}$, with $\beta=\pm i[2(1+\frac{1}{q-1})]^{1/2}$, which do not  have either any sort of  power-law decay for large $\mid y \mid$.

Then, it seems not unreasonable to infer that there are no acceptable functions $\epsilon (x)$ or that $0=\epsilon (x)=f_{2}(x)-f_{1}(x)$. Then, if  two functions  $f_{1}(x)$ and $f_{2}(x)$ have the same $q$-FT, it would follow that $f_{2}(x)=f_{1}(x)$, to  orders zeroth and first in $(q-1)$.

The above  arguments and uniqueness hold only to  first  order  in $(q-1)$ (which implied linearity in $f$). The analysis below, which allows for nonlinearities in $f$, will lead to different conclusions. 

2) We shall comment briefly about the classical inverse moment problem ($q=1$ case), i.e., whether the set $\{ \langle x^n \rangle_1\}$ corresponds to a unique normalized probability density $f_{1}(x)$ \cite{math1,math2,math3,math4}. Suppose that  two square-integrable functions  $f_{1}(x)$ and $f_{2}(x)$ ($-\infty<x<+\infty$) have the same moments $\{ \langle x^n \rangle_1\}$, all of which are finite. We also assume that the series $\sum_{0}^{+\infty}(n!)^{-1}(i\xi)^{n}\langle x^n \rangle_1$ converges (and that $\sum_{0}^{+\infty}$ and $\int_{-\infty}^{+\infty} \, 
dx$ can be interchanged) in a suitable range of $\xi$ values (*).
Then, $\epsilon (x)=f_{2}(x)-f_{1}(x)$ fulfills:
\be \label{appcmp}
\int_{-\infty}^{+\infty} \, 
dx \, x^n \,\epsilon (x)\, = 0\, \,\,\,\,(n=0,1,2,3,\ldots) \,.  
\ee
By subtracting the $q=1$ counterparts of Eq. (\ref{qFTsex})  for both  $f_{1}(x)$ and $f_{2}(x)$, it follows that: 
\be \label{appcmp1}
\int_{-\infty}^{+\infty} \, 
dx \, \exp i\xi x \,\epsilon (x)\, = 0\, \,\,\,\,(n=0,1,2,3,\ldots) \,.  
\ee
Since $\epsilon (x)$ is in the class of square-integrable functions,  Eq. (\ref{appcmp1}) implies that: $\epsilon (x)=0$.  Then,  no other normalized density $f_{2}(x)$ exists in  the  vicinity of $f_{1}(x)$, so that they both could have the same moments. 

Let us now replace the above condition (*) by:  the series $\sum_{0}^{+\infty}z^{-n}\langle x^n \rangle_1$ converges (and  $\sum_{0}^{+\infty}$ and $\int_{-\infty}^{+\infty} \, 
dx$ can be interchanged) in a suitable range of $\xi$ values (**). By summing a  geometric series, one has  for both  $f_{1}(x)$ and $f_{2}(x)$:
\be \label{appcmp11}
\sum_{0}^{+\infty}\frac{\langle x^n \rangle_1}{z^{n}}= \, 
  z\left[\int_{-\infty}^{+\infty} \, 
dx \, \frac{f_{1} (x)}{z- x }\right]=z\left[\int_{-\infty}^{+\infty} \, 
dx \, \frac{f_{2} (x)}{z- x }\right]
 \,\,\,.  
\ee
Then:
\be \label{appcmp12}
0= \, 
  z\left[\int_{-\infty}^{+\infty} \, 
dx \, \frac{\epsilon (x)}{z- x }\right]
 \,\,\,.
 \ee 
  The  structure of the right-hand-side of Eq. (\ref{appcmp12}) suggests that it  can be extended to an analytic function in the complex $z$-plane, except for a discontinuity across part of the    real $z$ axis. Such an analytic function has to vanish identically throughout the whole complex $z$-plane, by virtue of the uniqueness of analytic continuation. Then, its   discontinuity has  to vanish as well, so  that $\epsilon (x)=0$ for any $x$: uniqueness of the classical inverse moment problem under the assumed conditions.

At this point, we shall remind an example of non-uniqueness of the classical inverse moment problem, given  by Stieltjes (quoted by Chihara \cite{math3}): 
\be \label{Stie1}
\frac{1}{\pi^{1/2}\exp(1/4)}\int_{0}^{+\infty}dx  \, \exp(-(\ln x)^{2})
 \, x^n \,[1+C\sin(2\pi\ln x)]=\exp(\frac{(n+1)^{2}-1}{4})\equiv \langle x^n \rangle_1
\ee
which holds for any real constant  $\mid C\mid<1$. One could say that $f_{1}(x)$ corresponds to $C=0$ and  $f_{2}(x)$ to $C\neq 0$. Eq. (\ref{Stie1}) means that the probability distribution inside the integral gives  the same classical moments for any $C$! We shall recast (\ref{appcmp1}) into  the ($q=1$) Fourier Transform framework. Thus, we  can write formally:
\be \label{appcmp2}
\frac{1}{\pi^{1/2}\exp(1/4)}\int_{0}^{+\infty}dx  \, \exp(-(\ln x)^{2})
 \, \exp i\xi x  \,[1+C\sin(2\pi\ln x)]=\sum_{0}^{+\infty}\frac{(i\xi)^{n}}{n!}\exp(\frac{(n+1)^{2}-1}{4})
 \,.  
\ee
  This would seem to imply that   the  whole family of functions inside the integral in  (\ref{appcmp2}), as the real parameter $C$ varies (with  $\mid C\mid<1$), would have the same and unique Fourier transform! Such a conclusion is invalid, because  the series in the right-hand-side of (\ref{appcmp2}) diverges, precisely due to the growth  of $\exp(((n+1)^{2}-1)/4)$ with $n$. Then, in this case the condition that  $\sum_{0}^{+\infty}(n!)^{-1}(i\xi)^{n}\langle x^n \rangle_1$ converges is not fulfilled. Similarly, the formal counterpart of Eq. (\ref{appcmp11}) for the Stieltjes counterexample is: 
\be \label{appcmp22}
\frac{z}{\pi^{1/2}\exp(1/4)}\int_{0}^{+\infty}dx  \, \exp(-(\ln x)^{2})
 \,  \frac{1}{z- x } \,[1+C\sin(2\pi\ln x)]=\sum_{0}^{+\infty}\frac{1}{z^{n}}\exp(\frac{(n+1)^{2}-1}{4})
 \,.  
\ee
This would seem to imply that   the  whole family of functions inside the integral in  (\ref{appcmp22}), as the real parameter $C$ varies (with  $\mid C\mid<1$), would have the same and unique analytic continuation! Such a conclusion is again invalid, because  the series in the right-hand-side of (\ref{appcmp22}) diverges,  for the same reason as the one in (\ref{appcmp2}). The Stieltjes counterexample displays the crucial role  of the convergence conditions for those series, namely, either (*) or (**) for the classical inverse moment problem. Thus, one should expect that some convergence conditions for various series in the analysis of the inverse of the $q$-FT  will have to be imposed. 

Below, we shall be able to extend  Eqs. (\ref{appcmp11}) and  (\ref{appcmp12}) to the analysis of the inverse of the $q$-FT and, in Appendix B,  to the associated inverse moment problem. It seems apparent that  a related  analysis of the  inverse of the $q$-FT for $q\neq 1$ in   the $q=1$ Fourier Transform framework would meet difficulties.

3) Our starting point will now  be Eqs. (\ref{norm}), (\ref{teoremita1}) and (\ref{qFTsex}).  Suppose that  two functions  $f_{1}(x)$ and $f_{2}(x)$ have the same $q$-FT: $F_q[f_{1}](\xi)=F_q[f_{2}](\xi)$. Then, they have the same formal Taylor expansion, given in Eqs. (\ref{qFTsex}). We assume that, for some domain of $\xi$-values, the series in  Eq. (\ref{qFTsex})  and  the series 
\be \label{appA0}
  \sum_{n=0}^{+\infty}\,\frac{[(q-1)i\xi ]^{n}}{1+n(q-1)}\left[\int_{-\infty}^{+\infty} \, 
dx \, x^n \, [f_{j}(x)]^{1+n(q-1)}\right]
 \,\,.  
\ee
converge for both  $j=1,2$ (and that $\sum_{0}^{+\infty}$ and $\int_{-\infty}^{+\infty} \, 
dx$ can be interchanged). Such convergence conditions will play here a role similar to the condition (**) in item 2), for the classical inverse moment problem A comparison of the factor $i^n 
\left\{
\prod_{m=0}^{n-1}[1+m(q-1)]
\right\}/n!$ (present  in  Eq. (\ref{qFTsex})) with  $(1+n(q-1))^{-1} $ (in Eq. (\ref{appA0})) for large $n$ suggests that
if  Eq. (\ref{qFTsex}) converges,  then the convergence of Eq. (\ref{appA0}) would  not impose additional  essential restrictions.
Then: $ \left[ 
\frac{d^{(n)}F_q[f_{1}](\xi)}{d\xi^n} 
\right]_{\xi=0}=\left[ 
\frac{d^{(n)}F_q[f_{2}](\xi)}{d\xi^n} 
\right]_{\xi=0}$, not only for $n=1,2,...$ but also for $n=0$, by virtue of Eq. (\ref{norm}). By factoring  out $i^n 
\left\{
\prod_{m=0}^{n-1}[1+m(q-1)]
\right\}$ in  Eq. (\ref{teoremita1}), we get: 
\be \label{app2a}
\int_{-\infty}^{+\infty} \, 
dx \, x^n \, [f_{1}(x)]^{1+n(q-1)}\, = \,\int_{-\infty}^{+\infty} \, 
dx \, x^n \, [f_{2}(x)]^{1+n(q-1)}, \,\,\,\,(n=0,1,2,3,\ldots) \,.  
\ee

\noindent
One has:
\be \label{appA00}
  \sum_{n=0}^{+\infty}\,\frac{[(q-1)i\xi ]^{n}}{1+n(q-1)}\left[\int_{-\infty}^{+\infty} \, 
dx \, x^n \, [f_{1}(x)]^{1+n(q-1)}\right]=\sum_{n=0}^{+\infty}\,\frac{[(q-1)i\xi ]^{n}}{1+n(q-1)}\left[\int_{-\infty}^{+\infty} \, 
dx \, x^n \, [f_{2}(x)]^{1+n(q-1)}\right]
 \,\,.  
\ee

For fixed $f_{1}(x)$,  let  $f_{2}(x)-f_{1}(x)=\epsilon (x)$ be  small. 
The last equation  yields, by expanding into powers of $\epsilon (x)$ inside the integrals, keeping only  the first order in $\epsilon (x)$ and summing a  geometric series:

\be \label{H1}
0\, 
 = \sum_{n=0}^{+\infty}\,[(q-1)i\xi ]^{n}\left[\int_{-\infty}^{+\infty} \, 
dx \, x^n \, [f_{1}(x)]^{n(q-1)}\epsilon (x)\right]= 
 \int_{-\infty}^{+\infty} \, 
dx \, \frac{\epsilon (x)}{1-i\xi(q-1) xf_{1}(x)^{q-1} }\equiv H_{1}(\xi)\,
 \,\,.  
\ee

 Notice the formal similarity  between  $H_{1}(\xi)$ in  Eq. (\ref{H1}) and   Eq. (\ref{FqfUTS0}), except for the crucial exponent $1/(1-q)$ in the latter.
 $H_{1}(\xi)$ does  not coincide with $F_q[f](\xi)$, but it will provide a useful framework to discuss  local uniqueness versus non-uniqueness of  the $q$--Fourier Transform. If $z=[i\xi(q-1)]^{-1}$,  Eq. (\ref{H1}) can be recast as: 
\be \label{app2e}
G_{1}(z)=H_{1}(\xi) \, 
 = z\left[\int_{-\infty}^{+\infty} \, 
dx \, \frac{\epsilon (x)}{z- xf_{1}(x)^{q-1} }\right]
 \,\,\,.  
\ee
which is the $q\neq 1$ counterpart of   Eq. (\ref{appcmp12}).
The structure of Eq. (\ref{app2e}) suggests that $G_{1}(z)$, which vanishes by virtue of  Eqs. (\ref{H1}) and  (\ref{app2e}), can be extended to an analytic function in the complex $z$-plane, except for a discontinuity across the  real $z$ axis. Such an analytic function has to vanish identically throughout the whole complex $z$-plane, by virtue of the uniqueness of analytic continuation. If one could infer that $G_{1}(z)\equiv 0$  implies $\epsilon (x)\equiv 0$, that would indicate the local uniqueness of  the inverse to the  $q$-Fourier Transform, in a ``small'' set of  functions  which contains  $f_{1}(x)$. However, this will not be the case, as we shall see, due to the key structure $xf_{1}(x)^{q-1}$, genuine of the $q$-FT.

The following example will clarify the issue. We  turn to  the following class of normalizable  nonnegative probability densities  $f_{1}(x)$, $-\infty<x<+\infty$, with the following properties: 1) $f_{1}(-x)=f_{1}(x)$, 2) $f_{1}(0)$ is finite, 3) $f_{1}(x)$ decreases  monotonically in  $0<x<+\infty$, with $f_{1}(x)\rightarrow 0$ as  $x\rightarrow +\infty$. This class appears to include the Cauchy-Lorentz distribution.  As  $f_{1}(x)^{q-1}$  decreases  monotonically in  $0<x<+\infty$, it follows that $xf_{1}(x)^{q-1}$ vanishes at $x=0$, increases monotonically in $0<x<x_{0}$  and  decreases monotonically in $x_{0}<x<+\infty$. The value  $x_{0}$ is defined so that $ xf_{1}(x)^{q-1}$ takes on its  maximum ( denoted as $y_{0}>0$),  at $x=x_{0}$. Then, in  $0<x<+\infty$,  the function $xf_{1}(x)^{q-1}=y$ has two inverses, namely, $x_{1}(y)$ and $x_{2}(y)$, with $0\leq y \leq y_{0}$ ( $dx_{1}/dy>0$ and $dx_{2}/dy<0$).  One has:
\be \label{app2e11}
G_{1}(z)=G_{1,+}(z)+G_{1,-}(z) \,\ee
$G_{1,-}(z)$ and $G_{1,-}(z)$ are  the contributions from  $0<x<+\infty$ and  $0>x>-\infty$, respectively. As $G_{1}(z)\equiv 0$ and $G_{1,+}(z)$ and $G_{1,-}(z)$ have different domains of discontinuity, it follows that $G_{1,+}(z)=G_{1,-}(z)\equiv 0$. One has:
\be \label{app2e11+}
G_{1,+}(z) =  z\left[\int_{0}^{x_{0}} \, 
dx \, \frac{\epsilon (x)}{z- xf_{1}(x)^{q-1} }+\int_{x_{0}}^{+\infty} \, 
dx \, \frac{\epsilon (x)}{z- xf_{1}(x)^{q-1} }\right]\,\,\,.
\ee
  By performing the change of variables $x\rightarrow y$: 
\be\label{app2e13} G_{1,+}(z)\, 
 =  z\left[\int_{0}^{y_{0}} \, 
dy \, \frac{(dx_{1}/dy)\epsilon (x_{1}(y))+(dx_{2}/dy)\epsilon (x_{2}(y))}{z- y }\right] \,\,.  
\ee 
  Since  $G_{1,+}(z)\equiv 0$, it follows that $(dx_{1}/dy)\epsilon (x_{1}(y))+(dx_{2}/dy)\epsilon (x_{2}(y))=0$ for any $0<y<y_{0}$. But  this does not require that $\epsilon (x_{1}(y))=0$ and $\epsilon (x_{2}(y))=0$ separately, for any $0<y<y_{0}$, that is, there may be a cancellation between $\epsilon (x_{1}(y))$ and $\epsilon (x_{2}(y))$, due to the different signs of $dx_{1}/dy$ and $dx_{2}/dy$. That is, $\epsilon (x)$ is not forced to   vanish.
$G_{1,-}(z)$ can be  treated similarly and leads to the same conclusion.

 Then, there is not, in general,  local uniqueness of  the inverse to the  $q$-Fourier Transform. On the other hand,   local uniqueness of  the inverse to the  $q$-Fourier Transform holds indeed for restricted classes of functions: one of such classes is that formed by   $q$-Gaussians (with its specific constraints).

\section*{Appendix B: On the Characterization of  a  probability density  by all    escort mean values 
$\la x^n \ra_{q_n}$'s  together with all 
$\nu_{q_n}$'s}

We now investigate whether a probability density $f(x)$ can be uniquely 
 characterized by the set of all escort mean values 
$\la x^n \ra_{q_n}$ together with the  set of all associated quantities 
$\nu_{q_n}$.
Suppose that  two probability densities  $f_{1}(x)$ and $f_{2}(x)$ have the same $\la x^n \ra_{q_n}$ and the same $\nu_{q_n} [f_{1}]=\nu_{q_n} [f_{2}]$ ( Eqs. (\ref{norm}) and  Eqs. (\ref{escort4})) for all $n=0,1,2,....$. We continue to make the same assumptions on $f_{1}(x)$ and $f_{2}(x)$ as in item 3) of Appendix A, so that Eqs. (\ref{app2a}) and  (\ref{appA00}) hold. We shall add the following  condition: the series  
\be \label{appB0}
  \sum_{n=0}^{+\infty}\,\frac{[(q-1)i\xi ]^{n}}{1+n(q-1)}\left[\int_{-\infty}^{+\infty} \, 
dx \,  [f_{j}(x)]^{1+n(q-1)}\right]
 \,\,.  
\ee
converge for both  $j=1,2$ (and, again,  $\sum_{0}^{+\infty}$ and $\int_{-\infty}^{+\infty} \, 
dx$ can be interchanged) for some domain of $\xi$-values. As $\nu_{q_n} [f_{1}]=\nu_{q_n} [f_{2}]$, one has:
\be \label{appB00}
  \sum_{n=0}^{+\infty}\,\frac{[(q-1)i\xi ]^{n}}{1+n(q-1)}\left[\int_{-\infty}^{+\infty} \, 
dx \,  [f_{1}(x)]^{1+n(q-1)}\right]=\sum_{n=0}^{+\infty}\,\frac{[(q-1)i\xi ]^{n}}{1+n(q-1)}\left[\int_{-\infty}^{+\infty} \, 
dx \,  [f_{2}(x)]^{1+n(q-1)}\right]
 \,\,.  
\ee
Let  $f_{2}(x)-f_{1}(x)=\epsilon (x)$ is small, so that one recovers Eq.  (\ref{app2e}). 
Moreover,  by using  the same arguments as in 3)  in  Appendix A, with  $xf_{1}(x)^{q-1}$ replaced by $f_{1}(x)^{q-1}$, one gets: 
\be \label{app2B}
\int_{-\infty}^{+\infty} \, 
dx  \, [f_{1}(x)]^{n(q-1)}\epsilon (x)\, = 0\, \,\,\,\,(n=0,1,2,3,\ldots) \,.  
\ee
Moreover, Eq. (\ref{appB00}) yields:
\be \label{app2c} 
0=H_{2}(\xi) \, 
 = \sum_{n=0}^{+\infty}\,[(q-1)i\xi ]^{n}\left[\int_{-\infty}^{+\infty} \, 
dx  \, [f_{1}(x)]^{n(q-1)}\epsilon (x)\right]
  = z\left[\int_{-\infty}^{+\infty} \, 
dx \, \frac{\epsilon (x)}{z- f_{1}(x)^{q-1} }\right]=G_{2}(z)\,\,.  
\ee

 Both $G_{1}(z)$ and $G_{2}(z)$ can  be  extended to  analytic functions in the complex $z$-plane. On the other hand, they both have to   vanish identically throughout the whole complex $z$-plane. Then, the discontinuities of both  $G_{1}(z)$ and $G_{2}(z)$ across  the   real $z$ axis will provide two conditions on $\epsilon (x)$ and the question is whether  they suffice  to ensure $\epsilon (x)\equiv 0$.

We consider again  the same  class of normalizable  nonnegative probability densities  $f_{1}(x)$, $-\infty<x<+\infty$ as at the end of Appendix A,  which led to Eq. (\ref{app2e11}) and to the non-uniqueness to the inverse of the $q$-Fourier Transform. 
 We start with $G_{2}(z)$, which reads ($q>1$):
\be \label{app0} 
G_{2}(z)= z\left[\int_{0}^{+\infty} \, 
dx \, \frac{\epsilon (x)+\epsilon (-x)}{z- f_{1}(x)^{q-1} }\right]\,\,\,.  
\ee 
as $f_{1}(-x)^{q-1}=f_{1}(x)^{q-1}$. Since $f_{1}(x)^{q-1}$ is monotonic, $G_{2}(z)=0$ implies: $\epsilon (x)=-\epsilon (-x)$, to be used in what follows. We shall now consider: 
\be \label{app2e11B}
G_{1}(z)=G_{1,+}(z)+G_{1,-}(z) \, 
 = z\left[\int_{0}^{+\infty} \, 
dx \, \frac{\epsilon (x)}{z- xf_{1}(x)^{q-1} }-\int_{-\infty}^{0} \, 
dx \, \frac{\epsilon (-x)}{z- xf_{1}(x)^{q-1} }\right]\,\,\,.  
\ee
$G_{1,+}(z)$ and  $G_{1,-}(z)$ are  the first and second integrals in the right-hand-side of  Eq. (\ref{app2e11}), respectively. As the ranges  of discontinuity of $G_{1,+}(z)$ and $G_{1,-}(z)$ are  disjoint,  $G_{1,+}(z)\equiv 0$ and $G_{1,-}(z)\equiv 0$ follow.   
By performing the same change of variables $x\rightarrow y$ which led to Eq. (\ref{app2e13}): 
\be\label{app2e13BB} G_{1,+}(z)\, 
 =  z\left[\int_{0}^{y_{0}} \, 
dy \, \frac{(dx_{1}/dy)\epsilon (x_{1}(y))+(dx_{2}/dy)\epsilon (x_{2}(y))}{z- y }\right] \,\,.  
\ee 
As  $G_{1,+}(z)\equiv 0$, it follows that $(dx_{1}/dy)\epsilon (x_{1}(y))+(dx_{2}/dy)\epsilon (x_{2}(y))=0$ for any $0<y<y_{0}$. As there may be a cancellation between $\epsilon (x_{1}(y))$ and $\epsilon (x_{2}(y))$,  $\epsilon (x)$ is not forced to   vanish. The  consideration of $G_{1,-}(z)$ leads to a similar conclusion.

 Then,   a  probability density does not appear to be characterized uniquely by the set  of  all  its   escort mean values 
$\la x^n \ra_{q_n}$'s  together with all its
$\nu_{q_n}$'s, in general. However, as we already mentioned earlier, the convenient feature of uniqueness might occur for special classes of physically relevant densities, with special constraints.

\end{document}